\documentclass[conference]{IEEEtran}

\usepackage[acronym]{glossaries}
\usepackage{lettrine}
\usepackage{textcomp}
\usepackage{algorithm}
\usepackage[noend]{algpseudocode} 
\usepackage{amsmath}
\usepackage{amssymb}
\usepackage{siunitx}
\usepackage{graphicx} 
\usepackage{url}
\usepackage{xspace}
\usepackage{booktabs}
\usepackage{makecell}
\usepackage{tabularx}
\usepackage{array}

\begin{document}

\definecolor{ddgreen}{rgb}{0.00, 0.50, 0.00}
\newcommand{\gs}[1]{{\color{black}{#1}}}
\newcommand{\gsf}[1]{{\color{ddgreen}\textbf{\textit{[GAGAN: #1]}}}}

\newcommand{\xdna}{XDNA\texttrademark\xspace}
\newcommand{\xdnaone}{XDNA\texttrademark~1\xspace}
\newcommand{\xdnatwo}{XDNA\texttrademark~2\xspace}

\newcommand{\strix}{AMD Ryzen\texttrademark\xspace AI 9 HX 370\xspace}



\newacronym{AI}{AI}{artificial intelligence}
\newacronym{DL}{DL}{deep learning}
\newacronym{DNN}{DNN}{deep neural network}
\newacronym[longplural={application-specific integrated circuits}]{ASIC}{ASIC}{application-specific integrated circuit}
\newacronym[longplural={graphics processing units}]{GPU}{GPU}{graphics processing unit}
\newacronym{TPU}{TPU}{tensor processing unit}
\newacronym{CPU}{CPU}{central processing unit}
\newacronym[longplural={instruction set architectures}]{ISA}{ISA}{instruction set architecture}
\newacronym[plural=TCNs, firstplural={temporal convolutional neural networks (TCNs)}]{TCN}{TCN}{temporal convolutional neural network}
\newacronym[plural=CNNs, firstplural={convolutional neural networks (CNNs)}]{CNN}{CNN}{convolutional neural network}
\newacronym{MHSA}{MHSA}{multi-head self-attention}
\newacronym{MHA}{MHA}{multi-head attention}
\newacronym[longplural={reduced instruction set computers}]{RISC}{RISC}{reduced instruction set computer}
\newacronym{ARM}{ARM}{advanced RISC machine}
\newacronym{SSR}{SSR}{stream semantic register}
\newacronym{HW}{HW}{hardware}
\newacronym{WL}{WL}{workload}
\newacronym{ONNX}{ONNX}{open neural network exchange}
\newacronym{NAS}{NAS}{neural architecture search}
\newacronym{PULP}{PULP}{parallel ultra low power}
\newacronym{AXI}{AXI}{advanced extensible interface}
\newacronym{TCDM}{TCDM}{tightly coupled data memory}
\newacronym[longplural={micro-controller units}]{MCU}{MCU}{micro-controller unit}
\newacronym{ODL}{ODL}{on-device learning}
\newacronym{SGD}{SGD}{stochastic gradient descent}
\newacronym{FLOPS}{FLOPS}{floating point operations per second}
\newacronym{SNN}{SNN}{spiking neural network}
\newacronym{KWS}{KWS}{keyword spotting}
\newacronym{IIS}{IIS}{integrated systems laboratory}
\newacronym[longplural={large language models}]{LLM}{LLM}{large language model}
\newacronym[longplural={systems-on-chip}]{SoC}{SoC}{system-on-chip}
\newacronym{NLP}{NLP}{natural language processing}
\newacronym{SotA}{SotA}{state of the art}
\newacronym{CV}{CV}{computer vision}
\newacronym{MAC}{MAC}{multiply-accumulate}
\newacronym{IoT}{IoT}{internet of things}
\newacronym{SIMD}{SIMD}{single instruction multiple data}
\newacronym{PTQ}{PTQ}{post-training quantization}
\newacronym{QAT}{QAT}{quantization-aware training}
\newacronym{EEG}{EEG}{electroencephalogram}
\newacronym{RAW}{RAW}{read-after-write}
\newacronym{WRL}{WRL}{weight-reuse linear}
\newacronym{IRL}{IRL}{input-reuse linear}
\newacronym{LWT}{LWT}{layer-wise tiling}
\newacronym{DFT}{DFT}{depth-first tiling}
\newacronym{MQA}{MQA}{multi-query attention}
\newacronym{GQA}{GQA}{grouped-query attention}
\newacronym[longplural={vision transformers}]{ViT}{ViT}{vision transformer}
\newacronym{GELU}{GELU}{Gaussian error linear unit}
\newacronym{GEMM}{GEMM}{general matrix multiplication}
\newacronym{HPC}{HPC}{high-performance computing}
\newacronym[longplural={direct memory accesses}]{DMA}{DMA}{direct memory access}
\newacronym{PE}{PE}{processing element}
\newacronym{FPU}{FPU}{floating-point unit}
\newacronym{RAM}{RAM}{random-access memory}
\newacronym{SRAM}{SRAM}{static random-access memory}
\newacronym{DRAM}{DRAM}{dynamic random-access memory}
\newacronym{NE16}{NE16}{neural engine 16-channels}
\newacronym{FWSA}{FWSA}{fused-weight self-attention}
\newacronym{TQT}{TQT}{trained quantization threshold}
\newacronym{ReLU}{ReLU}{rectified linear unit}
\newacronym{ML}{ML}{machine learning}
\newacronym{SLM}{SLM}{small language model}
\newacronym{LM}{LM}{language model}
\newacronym{FC}{FC}{fully connected}
\newacronym{MLP}{MLP}{multi-layer perceptron}
\newacronym{OS}{OS}{operating system}
\newacronym{NPU}{NPU}{neural processing unit}
\newacronym{iGPU}{iGPU}{integrated graphics processing unit}
\newacronym{TOPS}{TOPS}{tera operations per second}
\newacronym{AIE}{AIE}{artificial intelligence engine}
\newacronym{VLIW}{VLIW}{very long instruction word}
\newacronym{NoC}{NoC}{network-on-chip}
\newacronym{MLIR}{MLIR}{multi-level intermediate representation}
\newacronym{IR}{IR}{intermediate representation}
\newacronym{CU}{CU}{compute unit}

\title{STEEL: Sparsity-Aware Fused Attention for Energy-Efficient Long-Sequence Inference \\on AMD's \xdna\ NPU}

\author{
    \IEEEauthorblockN{
    Victor J.B. Jung$^{\dagger}{}^{*}$, Gagandeep Singh$^*$, Joseph Melber$^*$, Kristof Denolf$^*$, Francesco Conti$^\ddagger$, Luca Benini$^{\dagger\ddagger}$
    }
    
    \IEEEauthorblockA{
    $^*$AMD Research and Advanced Development (RAD). \\
    $^\dagger$Integrated Systems Laboratory (IIS), ETH Zürich, Switzerland. \\ 
    $^\ddagger$ Department of Electrical, Electronic and Information Engineering (DEI), University of Bologna, Italy.
    }
}


\maketitle


\begin{abstract}

The growing adoption of \gls{LLM}-based agents within operating system workflows has increased the importance of energy-efficient inference on laptop-class \glspl{SoC}. While cloud offloading remains common, it introduces reliability and privacy concerns that are particularly problematic for agentic workloads.
Recent laptop \glspl{SoC}, therefore, incorporate \glspl{NPU} optimized for energy efficiency; however, effectively mapping attention mechanisms onto \glspl{NPU} remains challenging due to architectural diversity and explicit data-movement programming models.
In this work, we present STEEL, the first open-source implementation of FlashAttention targeting XDNA-like \glspl{NPU}. STEEL introduces a dataflow formulation of prefill attention, enabling efficient exploitation of spatial parallelism and on-chip memory. 
Furthermore, STEEL addresses the load imbalance induced by the causal mask by leveraging a sparsity-aware pipeline placement onto the \gls{NPU} array, reducing synchronization overhead and improving utilization. We evaluate STEEL on the \strix \gls{SoC} and compare its performance against optimized \gls{CPU} and \gls{GPU} implementations. Experimental results show that STEEL reduces energy consumption by an average of 9.17\,$\times$ and 1.75\,$\times$ relative to CPU and GPU baselines, respectively. On \xdnaone, STEEL achieves an average 9.6\,$\times$ latency reduction over the prior \gls{SotA}, and delivers a 22.8\,$\times$ speedup on average compared to a layer-by-layer attention implementation on \xdnatwo.

\end{abstract}

\glsresetall{}

\IEEEpeerreviewmaketitle

\section{Introduction}
\label{sec:introduction}

\lettrine[]{T}{he} \gs{increasing} integration of \gls{AI} agents into  \gs{core} operating system functions is a \gs{key driver in} the design of modern laptop \glspl{SoC}~\cite{mei_aios_2025}. These agents are typically \gs{implemented as} large Transformer-based \glspl{DNN} with several billion parameters~\cite{saad-falcon_intelligence_2025}. \gs{While such models enable powerful capabilities, their inference demands impose substantial computational and data-movement overhead, making them} inherently energy-intensive. \gs{This energy cost has emerged as a fundamental} bottleneck for embedded \gs{mobile} platforms\gs{, where power and thermal budgets are tightly constrained~\cite{saad-falcon_intelligence_2025}.}

\gs{As a result, most} \gls{LLM} inference is currently offloaded to \gs{data-center \glspl{GPU}. Although effective from a performance standpoint,} this centralized approach \gs{introduces challenges in agentic workflows, including increased latency, reduced} reliability, and \gs{heightened} privacy \gs{risks}~\cite{chen_survey_2025}.

To unlock the full potential of \gls{AI} agents at the edge,  recent laptop \glspl{SoC} \gs{integrate} \glspl{NPU} \gs{designed specifically for energy-efficient} inference~\cite{rico_amd_2024, lin_fastattention_2024, fei_nitro_2024}. \glspl{NPU} \gs{target the most computationally and energy expensive components of Transformer models, most notably the attention mechanism~\cite{yang_lserve_2025}.}
The prefill stage of Attention is a major contributor to \gs{inference} latency \gs{and energy consumption} at long-sequence length\gs{, a regime that is increasingly common in practical LLM deployments}~\cite{yang_lserve_2025}.
As a result, substantial effort has been made to optimize attention on \gs{both} commercial~\cite{dao_flashattention-2_2023} and academic hardware platforms~\cite{islamoglu_ita_2023}. The range of attention optimizations is broad, from algorithmic improvements like FlashAttention~\cite{dao_flashattention-2_2023} to hardware enhancements \gs{including} specialized non-linear units~\cite{prasad_pace_2025}.

\gs{\glspl{NPU} achieve high energy efficiency through spatial dataflow architectures and} explicit data-movement programming model\gs{s, which expose fine-grained control over computation and memory transfers~\cite{hunhoff_efficiency_2025}}.
Typical \gls{NPU} architectures, such as AMD's \xdna, are particularly well-suited for executing the prefill stage of language generation, as the prefill stage is primarily composed of matrix-to-matrix multiplication. Additionally, the prefill stage is a significant contributor to latency for large context requests~\cite{agrawal_taming_2024}, as is the case in agentic systems. 
While extensive prior work has focused on optimizing attention for \glspl{GPU}, comparatively few efforts target attention on \glspl{NPU}~\cite{lin_fastattention_2024}. Moreover, the substantial diversity among \glspl{NPU} architectures and programming models severely limits the portability of existing solutions.

In this work, we present STEEL, the first open-source implementation of FlashAttention on \xdnatwo \glspl{NPU}. 
STEEL proposes a dataflow formulation of the prefill attention distributed onto a three-stage pipeline of \gls{AIE} tiles. Through a careful distribution of work over the pipeline, optimized data-layout handling, and sparsity-aware pipeline placement, STEEL enhances the energy efficiency of the attention mechanism on edge platforms. 
\gs{This work makes the following key contributions:}

\begin{itemize}
    \item We propose STEEL, a dataflow formulation of FlashAttention for \xdna-like architectures. STEEL decomposes the FlashAttention algorithm to enable efficient distribution across the \gls{PE} array.
    \item We introduce a novel sparsity-aware pipeline placement \gs{technique} that mitigates workload distribution imbalance caused by the causal masking, hence reducing synchronization overhead. This placement achieves a 38\,\% \gs{latency} reduction compared to \gs{uniform} placement.
    \item We perform \gs{an in-depth} comparison of attention execution on the \strix \gls{SoC} across its \gls{NPU}, \gls{CPU}, and \gls{GPU}. In detail, we benchmark STEEL on the \xdnatwo \gls{NPU} and compare it against FlashAttention on 12 Zen5 \glspl{CPU} cores and the RDNA 3.5 \gls{GPU}.
    \item \gs{We demonstrate STEEL's portability across multiple \xdna-like architectures showing} that it outperforms the \gls{SotA} FlashAttention implementation from DATO~\cite{fang_dato_2025}.
\end{itemize}

Experimental evaluations of STEEL on \strix demonstrate an average energy consumption reduction of \textbf{9.17\,$\times$ and 1.75\,$\times$} compared to the \gls{CPU} and \gls{GPU}, respectively.
On \xdnaone, STEEL outperforms the previous \gls{SotA} implementation of flash-attention on \xdnaone~\cite{fang_dato_2025} by reducing the latency by \textbf{9.6\,$\times$} on average.
Additionally, compared to a layer-by-layer implementation of attention on \xdnatwo, STEEL provides an average 22.8\,$\times$ speedup.
The STEEL algorithm is open-source at \url{https://github.com/amd/iron}.


\vspace{-1em}

\section{Background}
\label{sec:background}

\subsection{Fused Attention Algorithms}
\label{sec:background:fused-attention}

Transformer blocks form the computational backbone of essentially all \glspl{LLM} and are used in a large fraction of modern large-scale \glspl{DNN}~\cite{grattafiori_llama_2024}. Within each transformer block, the \emph{attention mechanism}~\cite{vaswani_attention_2017} is responsible for much of the computational cost, memory footprint, and performance complexity.
As sequence lengths grow, attention often becomes the dominant contributor to inference latency and memory bottlenecks~\cite{yang_lserve_2025}.

The attention mechanism operates on three input tensors: queries $Q \in \mathbb{R}^{S_{q} \times d}$\, keys $K \in \mathbb{R}^{S_{kv} \times d}$, and values $V \in \mathbb{R}^{S_{kv} \times d}$, and produces an output tensor $O \in \mathbb{R}^{S_{q} \times d}$:

\begin{equation}
    A = \frac{QK^T}{\sqrt{d}}, \; O = softmax(A)V \; \text{where} \; A \in \mathbb{R}^{S_{q} \times S_{kv}} \gs{\label{section-back:eq:1}}
\end{equation}

Where $S_q$ and  $S_{kv}$ denote the query and key/value sequence lengths, respectively, and $d$ denotes the head dimension. The softmax function is applied independently across each row of A.

FlashAttention~\cite{dao_flashattention-2_2023} was introduced to eliminate the storage and latency overheads incurred by materializing $A$ and $softmax(A)$ in memory. Its key insight is to restructure the attention computation to eliminate explicit materialization of these intermediates entirely. FlashAttention achieves this by leveraging online softmax techniques~\cite{milakov_online_2018} and tiling the attention computation, allowing the attention score to be computed, normalized, and consumed incrementally. 
To preserve the correct normalization statistics across tiles, FlashAttention maintains two per-row statistics vectors, $\ell$ $m \in \mathbb{R}^{B_q}$, where $B_q$ is the user-selected block size for $Q$. %

\subsection{\xdna \;Software Stack}
\label{sec:background:xdna-sw-stack}

\begin{figure}[!t]
    \centering
    \includegraphics[width=\columnwidth]{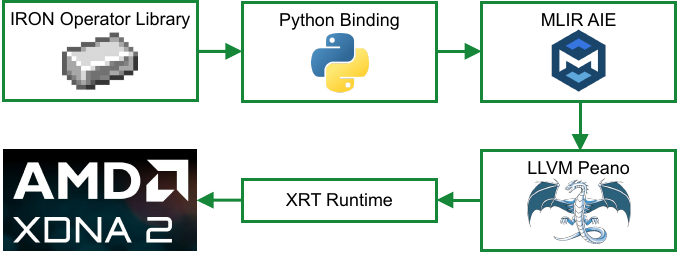}

    \vspace{-1em}
    \caption{Overview of the software stack used to program the \xdnatwo \gls{NPU}. The IRON library contains efficient \gls{ML} operators written in Python and using C++ kernels. The Python bindings are lowered to LLVM \gls{IR} by the MLIR-AIE compiler. The LLVM-AIE compiler generates binaries to run on the \gls{NPU}; the host-to-\gls{NPU} interactions are handled by the XRT runtime. The entire stack is composed of open-source tools.}
    \label{figure:XDNA-sw-stack}
    \vspace{-1em}
  
\end{figure}

To program the \xdnatwo \gls{NPU}, we use AMD’s open-source software stack (Fig.~\ref{figure:XDNA-sw-stack}). IRON provides compute primitives targeting both individual \gls{AIE} cores and the full \gls{NPU}. Single-core primitives (\textit{kernels}) are implemented in C++ using the \gls{AIE} API, while full-\gls{NPU} primitives (\textit{designs}) are implemented in Python via MLIR-\gls{AIE} bindings. Designs compose kernels and orchestrate data movement across AIE cores, Mem tiles, and \gls{DRAM}. For instance, the IRON \gls{GEMM} design distributes tiles across the array while invoking a C++ kernel for local computation.

The Python bindings define the dataflow over the \gls{PE} array and are lowered to the MLIR-\gls{AIE} dialect, optimized through transformation passes, and translated to LLVM \gls{IR}. This \gls{IR} is compiled by LLVM-AIE into binaries for the \gls{AIE} cores. Execution is managed by a C++ or Python runtime built on XRT or PyXRT, which either runs the operator on the \gls{NPU} and returns output tensors or executes validation test benches.

\subsection{\xdnatwo\ \gls{NPU} Architecture}
\label{sec:background:xdna2-arch}

\begin{figure}[b]
    \centering
    \includegraphics[width=\columnwidth]{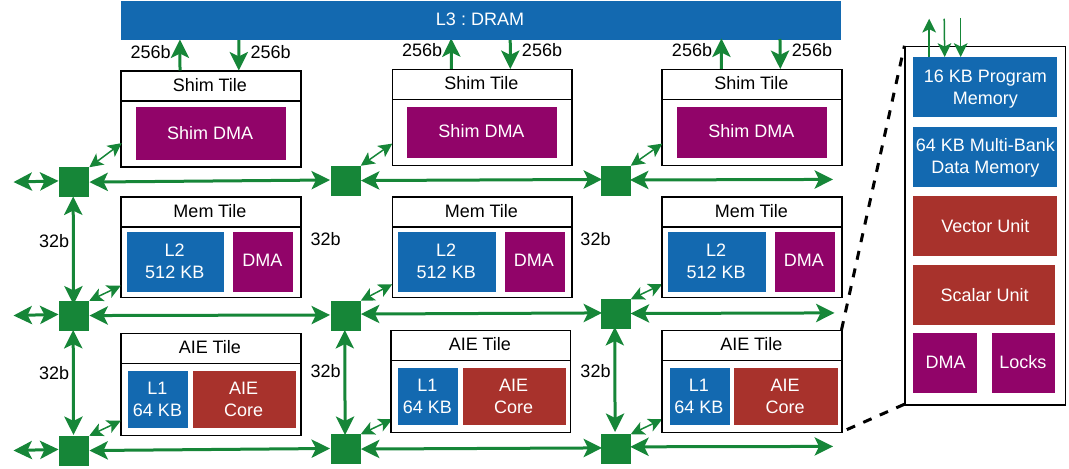}
  
    \caption{Overview of the \xdnatwo \gls{NPU}. \xdnatwo features three types of tiles interconnected via a \gls{NoC}. The Shim tiles feature high bandwidth \glspl{DMA} to move data in and out of the \gls{NPU}. The Mem tiles are intermediate large \SI{512}{\kilo\byte} buffers coupled with a \gls{DMA} engine. The \gls{AIE} tiles are \gls{VLIW} processors with a scalar and vector datapath; each \gls{AIE} core runs an independent program.}
    \label{figure:AIE-arch}
  
\end{figure}

\begin{table}[t]
\centering
\scriptsize
\renewcommand{\arraystretch}{1.15}
\setlength{\tabcolsep}{3pt}
\caption{Comparison of commercial \glspl{NPU} for laptop \glspl{SoC}.}
\label{table:npu-comparison}

\begin{tabular}{@{}m{0.20\columnwidth}m{0.17\columnwidth}m{0.17\columnwidth}m{0.18\columnwidth}m{0.17\columnwidth}@{}}
\toprule
& \centering Intel’s\\ AI Boost~\cite{fei_nitro_2024}
& \centering Qualcomm’s\\ Hexagon~\cite{codrescu_qualcomm_2013}
& \centering Huawei’s\\ Ascend\,310~\cite{dhar_ascend-cc_2024}
& \centering \textbf{AMD’s}\\ \textbf{XDNA\texttrademark 2~\cite{singh_sparta_2023}} \tabularnewline
\midrule
Open-Source Software Stack
& \centering Yes
& \centering No
& \centering Yes
& \centering Yes \tabularnewline

Spatial Dataflow \\Architecture
& \centering No
& \centering No
& \centering No
& \centering Yes \tabularnewline

Peak Throughput \\(TOPS)
& \centering 48
& \centering 45
& \centering 16
& \centering 50 \tabularnewline
\bottomrule
\end{tabular}

\vspace{-1em}
\end{table}

The \xdnatwo \gls{NPU}\cite{rico_amd_2024} adopts a two-dimensional spatial architecture composed of \gls{VLIW} processing units interconnected via a \gls{NoC} (Fig.~\ref{figure:AIE-arch}). It operates between \SI{1.3}{\giga\hertz} and \SI{1.8}{\giga\hertz} depending on power mode.
The architecture is organized into eight columns, each comprising a Shim tile, a Mem tile, and four \gls{AIE} tiles. The Shim tile connects the \gls{NoC} to \gls{DRAM} and provides a high-throughput \gls{DMA} engine supporting 4-D transfers over two 256-bit channels. The Mem tile integrates \SI{512}{\kilo\byte} of multi-bank interleaved memory and a \gls{DMA} engine with 4-D transfer support.
Computation is performed on \gls{AIE} tiles. Each \gls{AIE} core is a \gls{VLIW} processor issuing up to seven instructions across scalar and vector datapaths, enabling overlap of control and compute. The vector unit sustains 64 \gls{MAC} operations per cycle for bfloat16 inputs with fp32 accumulation. Each tile includes \SI{64}{\kilo\byte} of banked data memory, \SI{16}{\kilo\byte} of program memory, and an on-tile \gls{DMA} engine supporting 3-D transfers.

\section{Related Work}
\label{sec:related-work}

\subsection{Commercial \glspl{NPU}}

With the rise of early \glspl{DNN} for vision tasks such as \glspl{CNN}~\cite{redmon_you_2016}, specialized accelerators were integrated into edge processors to improve energy efficiency~\cite{moons_145_2017}. A similar trend is now emerging with the integration of \glspl{NPU} into portable \glspl{SoC} to accelerate \glspl{LLM}~\cite{saad-falcon_intelligence_2025}. However, commercially available \glspl{NPU} exhibit significant architectural diversity, making it important to contextualize the design space targeted by STEEL.
A primary comparison metric is peak throughput, typically reported in \gls{TOPS}. Table~\ref{table:npu-comparison} highlights large variation across platforms, reflecting differences in both scale and target workloads.
The spatial dataflow architecture of \xdnatwo (Sec.~\ref{sec:background:xdna2-arch}) differs fundamentally from other \glspl{NPU}. Its 32 \gls{AIE} cores execute independent programs and communicate through the \gls{NoC}, enabling fine-grained, operator-specific mappings across diverse \gls{ML} workloads. This flexibility expands the mapping design space, increasing the complexity of achieving efficient implementations.
Finally, practical deployment depends heavily on the maturity and accessibility of the software stack. Open-source stacks improve usability for both researchers and developers; Table~\ref{table:npu-comparison} shows that Hexagon remains the only platform relying on a closed software stack.

\subsection{Attention Mapping on \glspl{NPU}}

Optimizing the mapping of complex \gls{ML} operators, such as the attention mechanism, for a specific \gls{NPU} can lead to greatly improved performance compared to a naive approach~\cite{lin_fastattention_2024}. Efficient mappings still heavily rely on expert knowledge and are very time-consuming. Companies usually provide a library of hand-tuned mappings for their custom processors, the most well-known being BLAS~\cite{lawson_basic_1979}, CUDA, or ROCm. Several attempts have been made to optimize operator libraries~\cite{zhao_akg_2021}; however, due to the vast differences among hardware platforms, no uniform approach for automatically generating operator libraries has emerged yet.

Optimized mappings of the attention mechanism to \glspl{NPU} have recently been proposed. FastAttention~\cite{lin_fastattention_2024} maps attention onto the Ascend 310 \gls{NPU}. It introduces a multi-level tiling strategy and generates the causal mask on the fly, avoiding the need to store the mask and move it through the memory hierarchy.
STEEL shares some similarities with FastAttention in its masking strategy: we also find that generating the mask directly on the \gls{AIE} core is more efficient than fetching it from \gls{DRAM}. However, the Ascend \gls{NPU} architecture differs substantially. In FastAttention, the \gls{AI} cores do not communicate directly, and the attention computation is not distributed across multiple cores to form a balanced pipeline, unlike in STEEL. Moreover, FastAttention does not address the workload imbalance induced by the causal mask.
%
DATO~\cite{fang_dato_2025} presents a task-based programming model for executing several \gls{ML} operators on the \xdnaone \gls{NPU}, including attention. Although DATO implements a FlashAttention kernel, it does not discuss pipeline balancing nor the handling of sparsity induced by the attention mask.

\section{Methods}
\label{sec:methods}

\begin{algorithm}[t]

\caption{STEEL Algorithm}
\label{algo:steel}

\begin{algorithmic}[1]

\Procedure{First Stage}{$Q$, $K$} $\to (A, m, m`)$
    \For{$0 \leq i < T_q$}
        \State Acquire $Q_i$, $m$, $m`$
        \For{$0 \leq j < T_{kv}$}
            \State Acquire $K_j$, $A_{ij}$
            
            \State $A_{ij} \gets \text{matmul\_b\_transposed\_unswizzle}(Q_i, K_j)$ 
            \State $A_{ij} \gets \text{scale\_and\_mask}(A_{ij}, \frac{log2e}{\sqrt{d}})$
            \State $m` \gets \max(m, \mathrm{rowmax}(A_{ij}))$
            
            \State Release $K_j$, $A_{ij}$, $m$, $m`$
        \EndFor
        \State Release $Q_i$
    \EndFor
\EndProcedure

\Procedure{Second Stage}{$A$, $m$, $m`$} $\to (P, l, v)$
    \For{$0 \leq i < T_q$}
        \For{$0 \leq j < T_{kv}$}
            \State Acquire $A_{ij}$, $P_{ij}$, $m$, $m`$, $l$, $l`$, $v$
            
            \State $P_{ij} \gets e^{A_{ij}-m`}$
            \State $v \gets e^{m - m`}$
            \State $\ell \gets v \ell + rowsum(P_{ij})$
            \State $m \gets m` \,\text{and}\, l \gets l`$
            
            \State Release $A_{ij}$, $P_{ij}$, $m$, $m`$, $l$, $v$
        \EndFor
    \EndFor
\EndProcedure

\Procedure{Third Stage}{$P$, $V$, $l$, $v$} $\to (O)$
    \For{$0 \leq i < T_q$}
        \State Acquire $O_i$
        \For{$0 \leq j < T_{kv}$}
            \State Acquire $P_{ij}$, $V_j$, $l$, $v$
            
            \State $O_i \gets \text{scale\_swizzle}(O_i, v)$
            \State $O_i \gets \text{matmul}(P_{ij}, V_j)$ 
            
            \State Release $P_{ij}$, $V_j$, $\theta_1$
        \EndFor
        \State $O_i \gets \text{scale}\_swizzle(O_i, l^{-1})$
        \State Release $O_i$
    \EndFor
\EndProcedure

\end{algorithmic}
\end{algorithm}


\subsection{The STEEL Pipeline}
\label{sec:methods:steel-pipeline}

We design STEEL starting from a three-stage formulation of FlashAttention-2, where stages compute attention scores $A_{ij}$, apply online softmax, and update the $m$ and $\ell$ statistics while accumulating $P_{ij}V_j$ into $O_i$. Profiling this baseline reveals load imbalance across stages; we iteratively refine the decomposition to obtain a balanced pipeline.

Algorithm~\ref{algo:steel} shows the final formulation, where each procedure maps to one stage executed on a dedicated \gls{AIE} core. STEEL is implemented using the IRON Python Bindings API, which maps applications to the \xdnatwo \gls{NPU} and expresses inter-\gls{PE} communication via ObjectFIFO~\cite{hunhoff_efficiency_2025}. The \textit{First Stage} uses \textit{matmul\_b\_transposed\_unswizzle} to compute $A_{ij}$, while the \textit{Third Stage} uses \textit{matmul} to accumulate into $O_i$.

Efficient execution on \gls{AIE} cores requires tiles to match the vector-unit layout. Layout transformations can be performed either by the vector unit or via \gls{DMA}. Since matrix multiplication requires 4-D transfers and \gls{AIE}-local \gls{DMA} supports only 3-D, we use the Mem tile \gls{DMA}, which provides full 4-D support.

In the \textit{Third Stage}, $O_i$ is scaled by $e^{m-m’}$ and $l^{-1}$ using \textit{scale\_swizzle}. Each row is scaled independently while stored in a swizzled layout. To maximize vector utilization, we broadcast each scaling factor across 8 elements, concatenate them into a 64-element vector, and apply element-wise multiplication across 8 rows. This fully utilizes the \SI{512}{\bit} vector registers (64 $\times$ \SI{16}{\bit} elements).

To handle sparsity induced by the causal mask, each \gls{AIE} tile tracks the coordinates $(i, j)$ of tile $A_{ij}$, enabling three cases: (i) fully masked tiles are skipped, (ii) unmasked tiles are processed normally, and (iii) partially masked tiles apply the mask locally in the \textit{Second Stage}. This approach generalizes to other masking schemes, such as windowed attention in time-series models.

\subsection{Macroscale Data Movement}
\label{sec:methods:macroscale-data-movement}

\begin{figure}[t]
    \centering
    \includegraphics[width=\columnwidth]{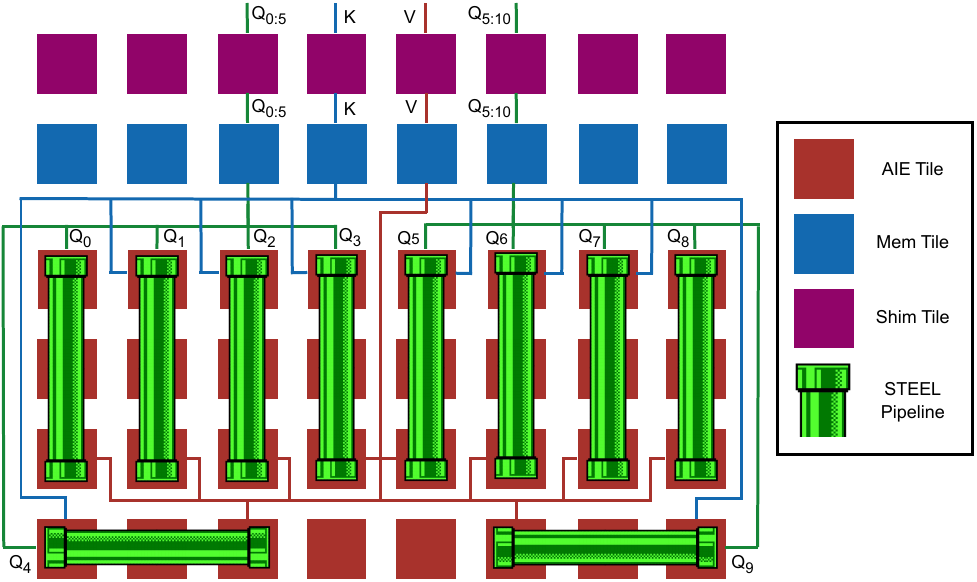}

    \vspace{-1em}
    \caption{Overview of the data movement between \gls{DRAM} and the STEEL pipelines. A chunk of 5 $Q$ tiles is brought down to the Mem tile, and then dispatched between 5 STEEL pipelines. Tiles of $K$ and $V$ are broadcast to each STEEL pipeline. Once a complete head of $K$ and $V$ has been broadcast, output tiles are sent back through the leftover Mem and Shim tiles.}
    \label{figure:pipeline-placement}
  
\end{figure}

Subsection~\ref{sec:methods:steel-pipeline} describes the intra-pipeline algorithm and data movement; mapping STEEL to hardware additionally requires placing pipelines on the \gls{PE} array and orchestrating data transfers between \gls{DRAM} and the \gls{NPU}. Figure~\ref{figure:pipeline-placement} shows the pipeline placement and the movement of $Q$, $K$, and $V$ tiles. We use the IRON \textit{distribute} primitive to assign one $Q$ tile per pipeline, \textit{join} to collect $O$ tiles, and broadcast $K$ and $V$ to all pipelines.

Memory port availability is a key constraint on \xdnatwo. Each Mem tile provides six ports, limiting the number of concurrent ObjectFIFOs per tile. With 10 pipelines, we therefore use two Mem tiles to distribute each wave of 10 $Q$ tiles.
To swizzle the $P_{ij}$ tiles between the second and third stages of the STEEL pipeline, we use the Mem tile \gls{DMA} engine. Consequently, each STEEL pipeline consumes four Mem tile ports: one for $Q$, one for $O$, and two for swizzling $P$. In addition, the pipelines collectively share two Mem tile ports for broadcasting $K$ and $V$. Overall, the 10 STEEL pipelines use 42 Mem tile ports out of 48.

\subsection{Sparsity-Aware Pipeline Placement}
\label{sec:methods:sparsity-aware-placement}

Broadcasting $K$ and $V$ is required to stay within the Mem tile port budget (Subsection~\ref{sec:methods:macroscale-data-movement}), but introduces a synchronization constraint: a broadcast can start only when all consumers are ready. While this is trivial under uniform workloads, the causal mask in \glspl{LM} creates load imbalance across STEEL pipelines. Although double buffering mitigates this effect, we find that explicitly balancing sparsity across pipeline groups significantly improves runtime.

\begin{figure}[b]
    \centering
    \includegraphics[width=\columnwidth]{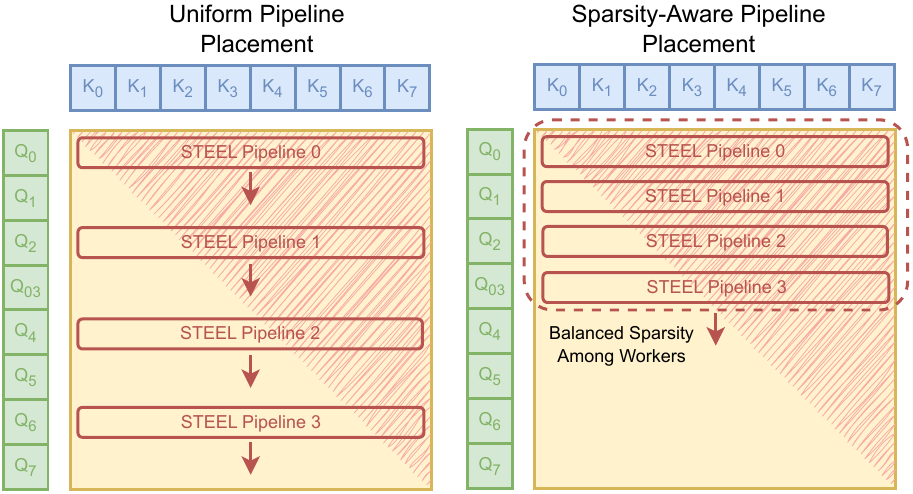}

    \vspace{-1em}
    \caption{Description of two pipeline placement strategies over the attention matrix $A$. On the left side, the STEEL pipelines are placed uniformly on $A$, resulting in an imbalance of the sparsity amount in the pipeline group and causing stalls. On the right side, we show our sparsity-aware pipeline placement where the sparsity between STEEL pipelines within a group is close.}
    \label{figure:sparsity-aware-placement}
  
\end{figure}

Figure~\ref{figure:sparsity-aware-placement} compares two placement strategies over the attention matrix $A$. Uniform placement assigns contiguous $T_q$ chunks to pipelines, resulting in uneven sparsity within a group. We instead propose a sparsity-aware placement that distributes consecutive rows across pipelines to equalize sparsity, yielding a 38,\% speedup.

\vspace{-2em}

\section{Results}
\label{sec:results}

In this section, we describe our evaluation setup and present an extensive benchmark of STEEL. We begin by comparing STEEL against a standard layer-by-layer attention implementation on \xdnaone. We then benchmark STEEL against the \gls{SotA} FlashAttention implementation provided by DATO~\cite{fang_dato_2025}, also on \xdnaone. Finally, in subsection~\ref{sec:results:strix-benchmark}, we benchmark the fused attention across the processing units available in the \strix \gls{SoC}. \strix integrates 16 Zen5 \glspl{CPU}, an RDNA 3.5 \gls{GPU}, and the \xdnatwo \gls{NPU}.

\subsection{Evaluation Setup}
\label{sec:results:eval-setup}

Results from subsections~\ref{sec:results:attention-impl-benchmark} and~\ref{sec:results:strix-benchmark} were collected on the \strix \gls{SoC}, a TSMC \SI{4}{\nano\meter} single-die platform with 12 Zen5 cores (24 threads) up to \SI{5.1}{\giga\hertz}. The RDNA 3.5 \gls{GPU} comprises 16 \glspl{CU} (1024 shaders at \SI{2900}{\mega\hertz}, GFX1150 ISA) and shares \SI{32}{\giga\byte} \gls{DRAM} with the \gls{CPU} and \gls{NPU}. The system also integrates the \xdnatwo \gls{NPU} (see Subsection~\ref{sec:background:xdna2-arch}).

CPU and GPU benchmarks use the TorchLib C++ frontend from PyTorch 2.1~\cite{paszke_pytorch_2019}, leveraging ROCm 6.4 and the HIP runtime, with \textit{scaled\_dot\_product\_attention} executed via the FlashAttention backend. The NPU is configured in turbo mode using XRT (\SI{1.8}{\giga\hertz}). Latency results are averaged over 50 iterations after 25 warm-ups.
Power is measured using AMD’s AGT tool at a \SI{50}{\milli\second} sampling rate over 100 iterations, reporting average consumption.

\subsection{Attention Implementation Benchmark}
\label{sec:results:attention-impl-benchmark}

\begin{figure}[t]
    \centering
    \includegraphics[width=\columnwidth]{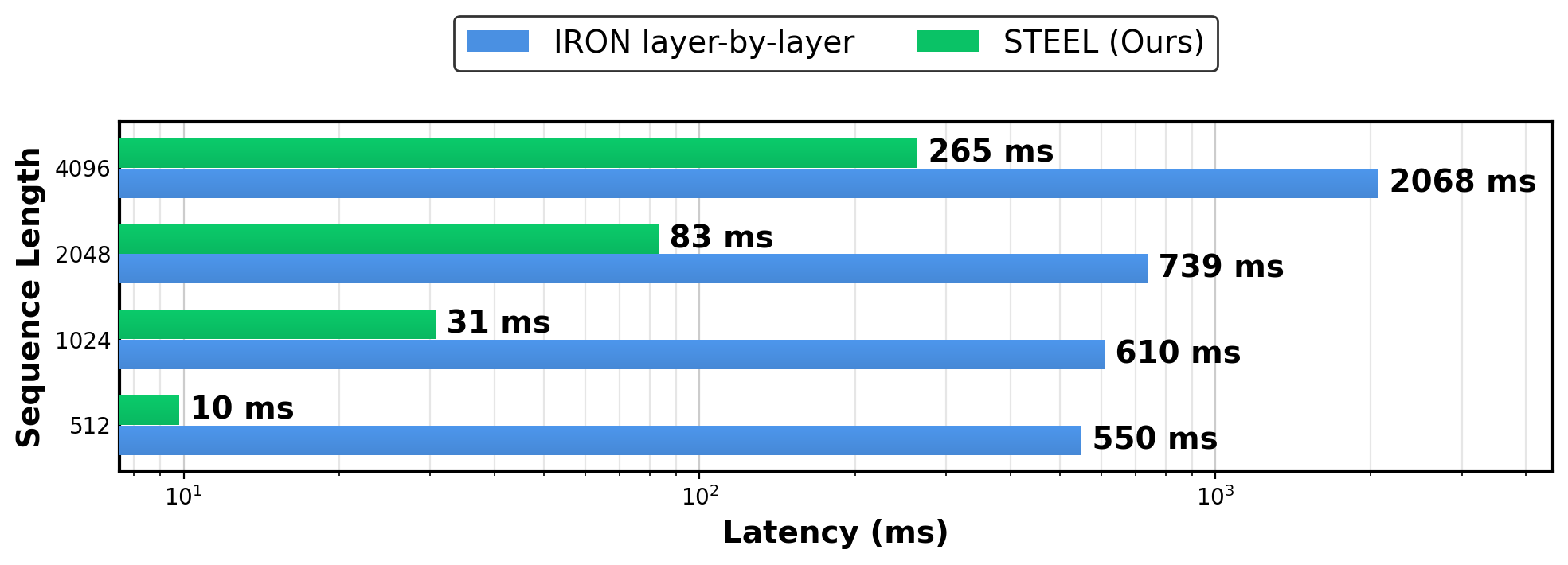}

    \vspace{-1em}
    \caption{Benchmark of the layer-by-layer attention against the STEEL's fused-attention on \xdnaone. The attention configuration is from BERT, with 12 heads and a head dimension of 64.}
    \label{figure:steel-iron-benchmark}
    \vspace{-2em}
    
\end{figure}

Figure~\ref{figure:steel-iron-benchmark} compares STEEL with a layer-by-layer attention implementation built from IRON operators. STEEL achieves a 22.8$\times$ speedup on average, demonstrating the benefit of fused attention. This gain stems from reduced overhead: STEEL loads a single design onto the \gls{NPU}, whereas the baseline requires separate \gls{GEMM}, Softmax, and Scale kernels, incurring context-switch costs. Additionally, the baseline transfers intermediate tensors $A$ and $P$ between the \gls{NPU} and \gls{DRAM}, while STEEL avoids this by not materializing them off-chip.

\begin{figure}[t]
    \centering
    \includegraphics[width=\columnwidth]{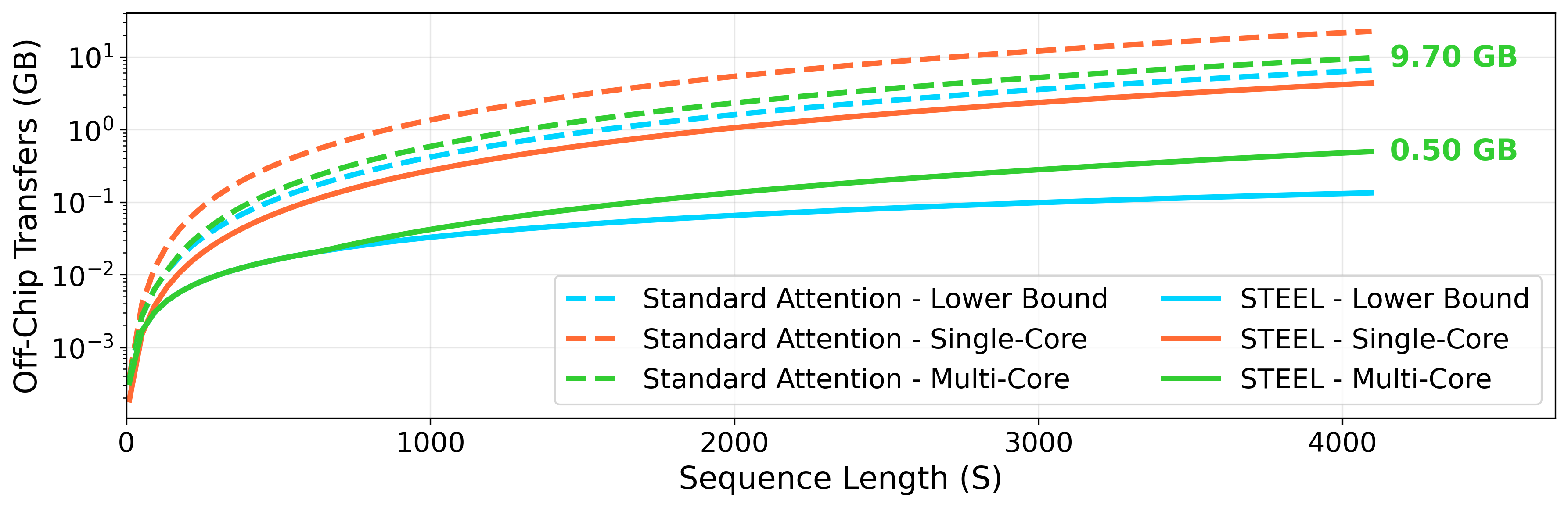}
  
    \caption{Off-chip transfer volume required to execute attention on \xdnatwo across several implementations as a function of sequence length. Dashed curves denote standard layer-by-layer attention implementations, whereas solid curves correspond to STEEL. Blue curves indicate the theoretical lower bound for each implementation and assume an infinite on-chip buffer capacity. \emph{Single-core} uses one \gls{AIE} core, while \emph{multi-core} uses the full \xdnatwo array.}
    \label{figure:memory-transfer-model}
  
\end{figure}

Figure~\ref{figure:memory-transfer-model} compares off-chip data movement for layer-by-layer (standard) and fused attention across sequence lengths. While individual IRON operators are optimized for locality, fusion further improves it. The model accounts for the IRON \gls{GEMM} dataflow and STEEL’s dataflow to estimate transfer volume. Blue curves indicate the theoretical lower bound; in practice, limited on-chip buffering leads to higher realized traffic.
Figure~\ref{figure:memory-transfer-model} also shows how off-chip traffic scales with core count. Multi-core execution on the full \xdnatwo \gls{NPU} reduces transfers due to increased on-chip buffering and broadcast reuse. For a sequence length of 4096, STEEL achieves a 19.4$\times$ reduction, from \SI{9.7}{\giga\byte} to \SI{0.5}{\giga\byte}.

\subsection{Comparison with State-of-the-Art on \gls{NPU}}
\label{sec::results:npu-sota-benchmark}

\begin{figure}[t]
    \centering
    \includegraphics[width=\columnwidth]{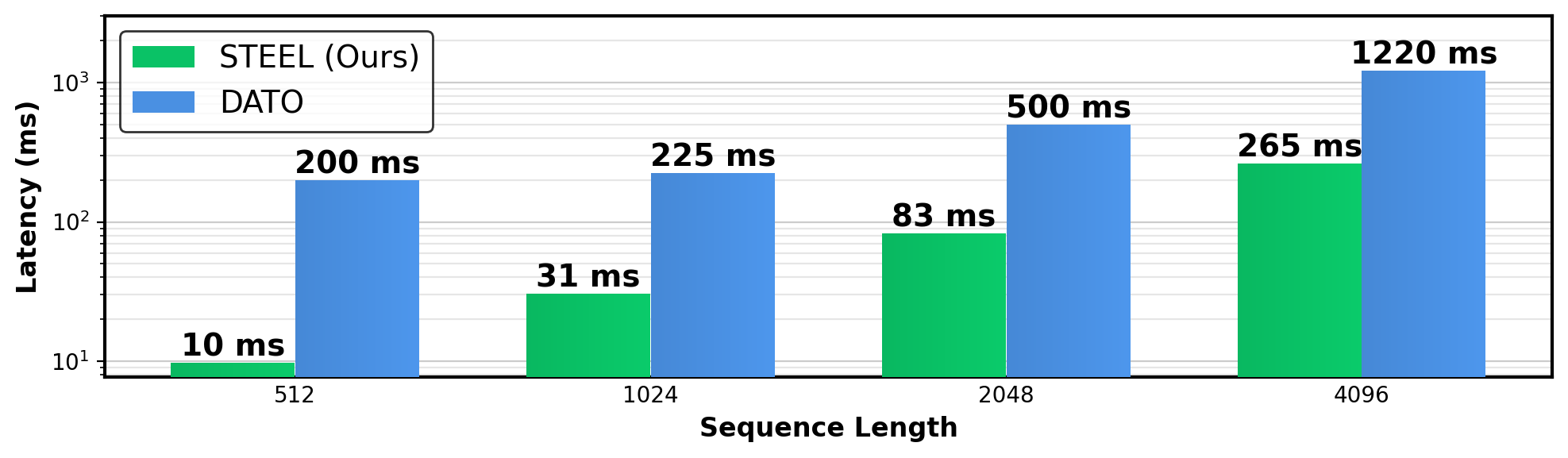}

    \vspace{-1em}
    \caption{Benchmark of STEEL against DATO~\cite{fang_dato_2025} on \xdnaone for several sequence lengths. The attention configuration is from BERT, with 12 heads and a head dimension of 64.}
    \label{figure:steel-dato-benchmark}
  
\end{figure}

To the best of the authors’ knowledge, DATO~\cite{fang_dato_2025} is the only published work that reports FlashAttention latency on \xdna \glspl{NPU}. However, DATO targets the \xdnaone \gls{NPU}. To show that STEEL’s performance is not specific to a single \gls{NPU} generation, we port STEEL to \xdnaone.
Whereas \xdnatwo provides eight columns of \glspl{PE}, \xdnaone provides five. Accordingly, we deploy one STEEL pipeline per column on the first three \gls{AIE} cores. We then place an additional STEEL pipeline on the last \gls{AIE} core of each of the first three columns, as illustrated in Figure~\ref{figure:pipeline-placement}. Finally, \xdnaone \gls{AIE} cores do not include a dedicated exponentiation unit; instead, we implement exponentiation via a lookup table.
We benchmark against DATO in an application-relevant setting by selecting the BERT attention configuration, which uses 12 heads with a head dimension of 64. In Figure~\ref{figure:steel-dato-benchmark}, we observe that STEEL consistently outperforms DATO across all sequence lengths. On average, STEEL accelerates attention by 9.6$\times$ relative to DATO.
This speedup primarily stems from how DATO partitions FlashAttention into four stages, which introduces a fundamental load imbalance in the pipeline. In particular, DATO performs output rescaling in the final stage. This rescaling is an element-wise multiplication between the precomputed factor $e^{m_{j-1}-m_j}$ and the current output tile $O_j$. Consequently, the fourth stage performs only $B_q \cdot B_{kv}$ \glspl{MAC}, which is substantially less than the $B_q \cdot B_{kv} \cdot d$ work required for the \gls{GEMM} in the first stage. For the BERT head dimension $d=64$, the last stage therefore performs 64$\times$ less computation than the first stage, resulting in a suboptimal mapping.
Additionally, we were unable to benchmark DATO for sequence lengths greater than 4096 due to what appears to be an exponential increase in compilation time.

\subsection{Attention Benchmark on \strix}
\label{sec:results:strix-benchmark}

\begin{figure}[t]
    \centering
    \includegraphics[width=\columnwidth]{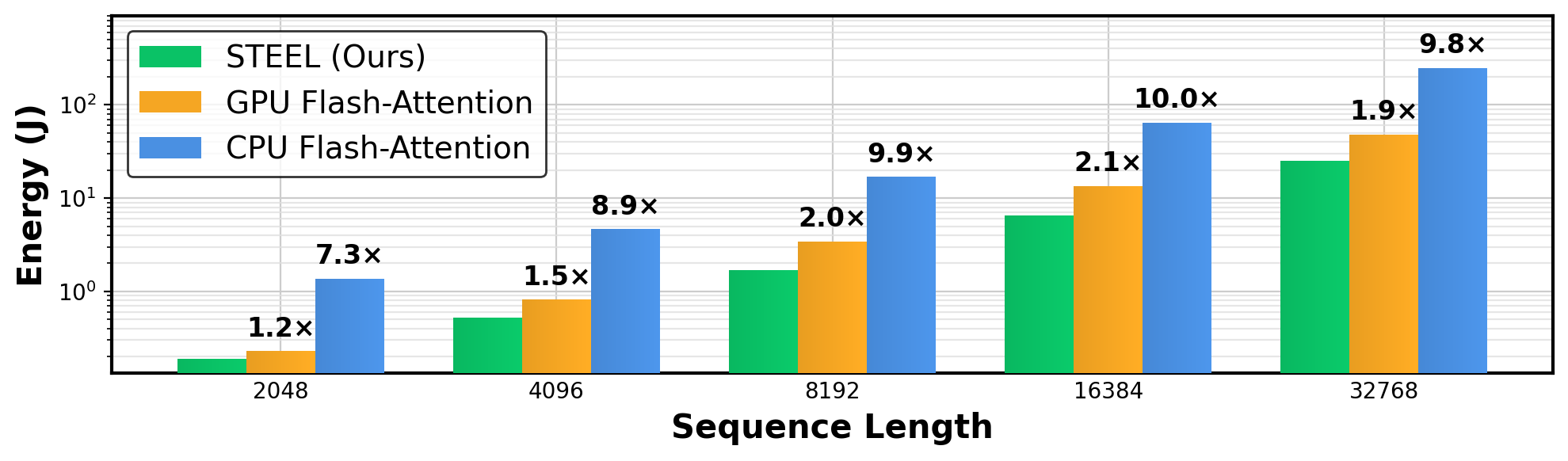}
  
    \caption{Benchmark of the attention energy efficiency on AMD Ryzen\texttrademark\xspace AI 9 HX 370 for various sequence lengths. The attention’s configuration is from Llama3.1-1B~\cite{grattafiori_llama_2024} with 32 heads and a head dimension of 64.}
    \label{figure:npu-cpu-gpu-benchmark}
    \vspace{-1em}
  
\end{figure}

\strix is a heterogeneous \gls{SoC} in which the \xdnatwo \gls{NPU} is the preferred compute engine for low-power \gls{DNN} inference. 
%
To verify that the \gls{NPU} is the most appropriate engine for attention inference, we benchmark optimized attention kernels across the three compute engines available in \strix, as shown in Figure~\ref{figure:npu-cpu-gpu-benchmark}: the \xdnatwo \gls{NPU}, the RDNA 3.5 \gls{GPU}, and the Zen5 \gls{CPU}.
To emulate an application-representative workload, we adopt the attention dimensions of Llama3.1-1B~\cite{grattafiori_llama_2024}. In this model, attention uses 32 heads with a head dimension of 64, and the maximum supported context length is 128k tokens. We sweep the sequence length from 2048 to 32768 to capture a common usage regime in which the model processes substantial contextual information (e.g., a codebase).
On average, STEEL reduces energy consumption by 9.17$\times$ and 1.75$\times$ relative to the \gls{CPU} and \gls{GPU}, respectively.
We further observe that STEEL’s energy-efficiency advantage over both the \gls{CPU} and the \gls{GPU} grows with sequence length, most notably between 2048 and 8192. This trend shows that the STEEL pipeline reaches a warmer steady state and attains near-peak throughput at approximately a sequence length of 8192.

\vspace{-1em}
\section{Conclusion}
\label{sec:conclusion}

We presented STEEL, a dataflow formulation of the FlashAttention algorithm targeting the \xdna \gls{NPU} family. STEEL carefully balances the workload across a three-stage pipeline of \gls{AIE} cores and uses a sparsity-aware pipeline placement to mitigate the workload distribution imbalance induced by causal masking. 
On \xdnaone, STEEL outperforms the previous \gls{SotA} implementation of flash-attention on \xdnaone~\cite{fang_dato_2025} by reducing the latency by 9.6\,$\times$ on average.
On the \strix \gls{SoC}, STEEL reduces energy consumption by 9.17$\times$ and 1.75$\times$ relative to the \gls{CPU} and \gls{GPU}, respectively.

\vspace{-0.5em}
\section*{Acknowledgment}

This work has received funding from the Swiss State Secretariat for Education, Research, and Innovation (SERI) under the SwissChips initiative.

\bibliographystyle{IEEEtran}
\bibliography{references}

\end{document}